\documentclass[12pt]{article}
\usepackage{graphicx}
\usepackage{xspace}

\newcommand\pubnumber{SNSN-323-63}
\newcommand\pubdate{March 30, 2021}

\def\institute{CERN,  Geneva,  Switzerland}

\def\Title#1{\begin{center} {\Large #1 } \end{center}}
\def\Author#1{\begin{center}{ \sc #1} \end{center}}
\def\Address#1{\begin{center}{ \it #1} \end{center}}

\newcommand\pubblock{\rightline{\begin{tabular}{l} \pubnumber\\
         \pubdate  \end{tabular}}}
\newenvironment{Abstract}{\begin{quotation}  }{\end{quotation}}
\newenvironment{Presented}{\begin{quotation} \begin{center} 
             PRESENTED AT\end{center}\bigskip 
      \begin{center}\begin{large}}{\end{large}\end{center} \end{quotation}}

\def\beq{\begin{equation}}
\def\eeq#1{\label{#1}\end{equation}}
\def\eeqn{\end{equation}}

\def\beqa{\begin{eqnarray}}
\def\eeqa#1{\label{#1}\end{eqnarray}}
\def\eeqan{\end{eqnarray}}

\let\bar=\overbar

\def\Dslash{\not{\hbox{\kern-4pt $D$}}}
\def\dslash{\not{\hbox{\kern-2pt $\del$}}}

\def\msb{{\bar{\ssstyle M \kern -1pt S}}}

\newcommand{\ttbar}{\ensuremath{\mathrm{t\bar{t}}}\xspace}
\newcommand{\mt}{\ensuremath{m_\mathrm{t}}\xspace}
\newcommand{\mtmc}{\ensuremath{\mt^{\mathrm{MC}}}\xspace}\newcommand{\mtmt}{\ensuremath{\mt(\mt)}\xspace}
\newcommand{\mtp}{\ensuremath{m_\mathrm{t}^{\mathrm{pole}}}\xspace}

\newcommand{\as}{\ensuremath{\alpha_\mathrm{S}}\xspace}
\newcommand{\msbar}{\ensuremath{\mathrm{\overline{MS}}}\xspace}
\newcommand{\TeV}{\ensuremath{\,\mathrm{TeV}}\xspace}
\newcommand{\GeV}{\ensuremath{\,\mathrm{GeV}}\xspace}

\newcommand{\fbinv}{\ensuremath{\;\mathrm{fb}^{-1}}\xspace}
\newcommand{\stt}{\ensuremath{\sigma_{\ttbar}}\xspace}
\newcommand{\roots}{\ensuremath{\sqrt{s}}}
\newcommand{\emu}{$\mathrm{e^{\mp}}\mu^{\pm}$\xspace}

\begin{document}
\begin{titlepage}
\pubblock

\vfill
\Title{Measuring standard model parameters using top-quark \\ cross sections in ATLAS and CMS}
\vfill
\Author{Matteo M.  Defranchis \\ on behalf of the ATLAS and CMS Collaborations}
\Address{\institute}
\vfill
\begin{Abstract}
Theoretical predictions of top-quark cross sections depend on the values of the parameters of the quantum chromodynamics Lagrangian, such as the strong coupling constant and the mass of the top quark, but also on parameters of the electroweak sector of the standard model and parton distribution functions.  Comparisons between state-of-the-art calculations and recent measurements at the ATLAS and CMS experiments at the CERN LHC allow for the precise determination of these parameters. In this proceeding, the most recent results by the two collaborations are summarized and discussed.
\end{Abstract}
\vfill
\begin{Presented}
$13^\mathrm{th}$ International Workshop on Top Quark Physics\\
Durham, UK (videoconference), 14--18 September, 2020
\end{Presented}
\vfill
\end{titlepage}
\def\thefootnote{\fnsymbol{footnote}}
\setcounter{footnote}{0}

\section{Introduction}

In proton-proton (pp) collisions at the CERN LHC, top quarks are mainly produced in quark-antiquark pairs (\ttbar) via the mechanism of gluon fusion. The \ttbar production cross section can be calculated in perturbative quantum chromodynamics (QCD), and is currently know up to the next-to-next-to-leading order (NNLO) in perturbation theory, including next-to-next-to-leading logarithmic (NNLL) corrections~\cite{Czakon:2011xx}.  Differential and multi-differential calculations at NNLO are also becoming available for an increasing number of variables~\cite{Catani:2019hip}.  The predicted cross sections depend on the values of fundamental parameters of the QCD Lagrangian such as the mass of the top quark, \mt, and the strong coupling constant, \as, but also on the parton distribution functions (PDFs) of the proton.  Therefore,  measurements of the inclusive and differential \ttbar production cross sections can be used to extract the values of \mt, \as, and to probe the proton PDFs.  In addition, this method allows the value of \mt to be determined in a well-defined renormalization scheme (on-shell, \msbar), overcoming the difficulties related to the interpretation of the results of direct \mt measurements. These are often referred to as top quark Monte Carlo (MC) mass (\mtmc) measurements, as the results depends on the details of the MC simulation~\cite{Butenschoen:2016lpz}.  However, it has to be considered that cross section measurements have a residual dependence on the value of \mtmc due to acceptance corrections, which has to be taken into account. Furthermore, the value of \mtmc also affects the efficiency of most kinematic reconstruction methods, which are often used in differential cross section measurements.

In this proceeding, the most recent results obtained by the ATLAS~\cite{Aad:2008zzm} and CMS~\cite{Chatrchyan:2008aa} Collaborations are summarized and discussed. The results are obtained using pp collision data at different centre-of-mass energies. In certain cases, the precision of the measurements outperform the one of the corresponding fixed-order calculations, making theoretical uncertainties a limiting factor.

\section{Extraction of QCD parameters}

\begin{figure}
\centering
\includegraphics[width=.49\textwidth]{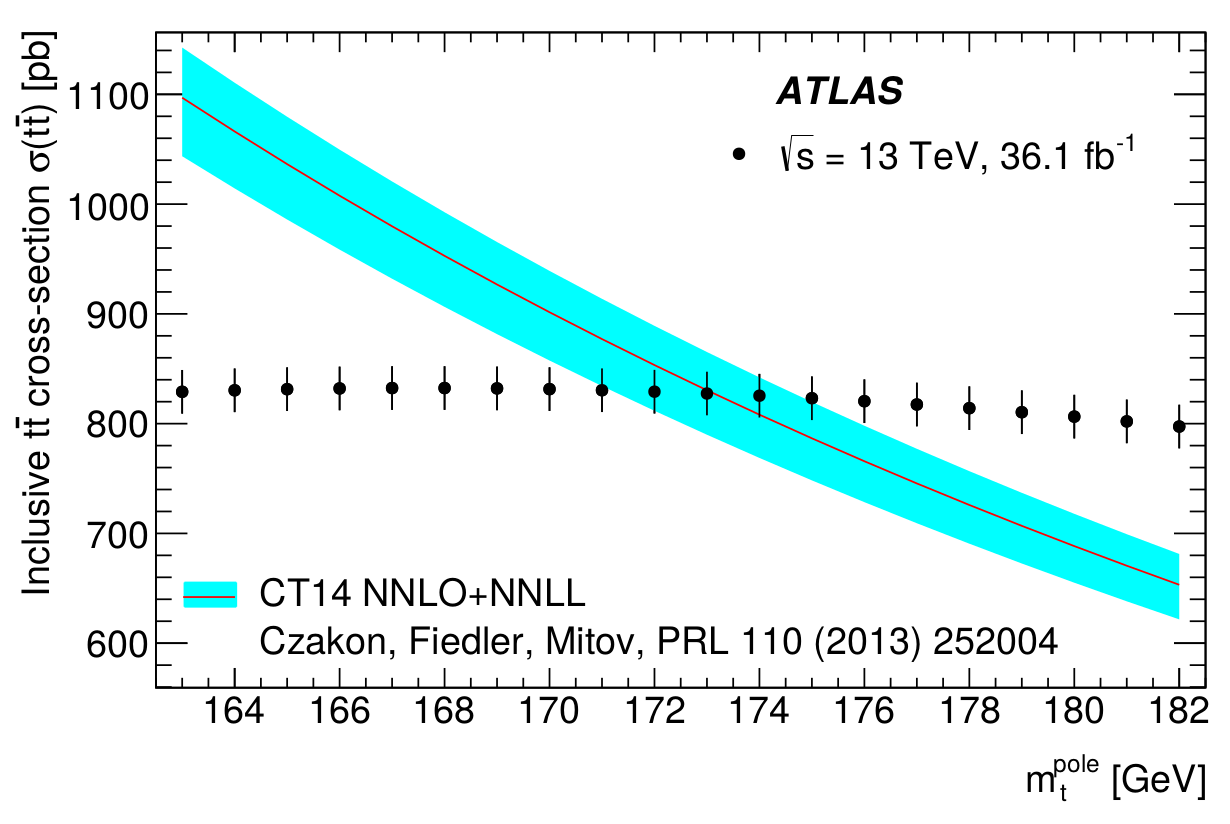}
\includegraphics[width=.49\textwidth]{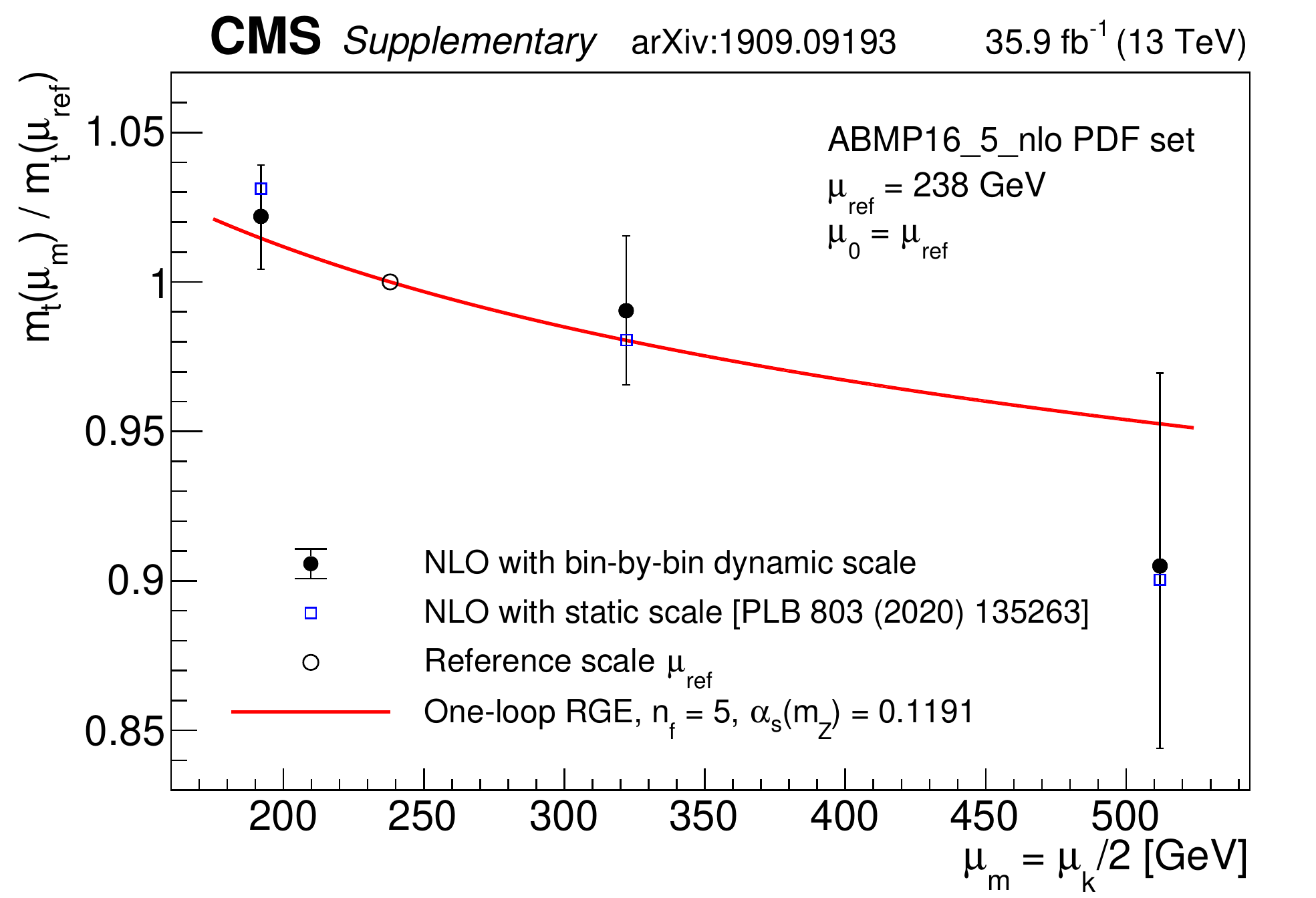}
\caption{Left: Measured (markers) and predicted (band) \stt at $\roots = 13\TeV$ as a function of \mtp, assuming $\mtmc = \mtp$.  These results are used to extract the value of \mtp from the measured \stt~\cite{Aad:2019hzw}. Right: Running of the top quark mass as a function of the scale $\mu_m$, defined as half of the invariant mass of the \ttbar system, using theoretical calculations with fixed or dynamic scales~\cite{Sirunyan:2019jyn}.}
\label{fig}
\end{figure}

The most precise measurement to date of the inclusive \ttbar production cross section, \stt, has been obtained by the ATLAS Collaboration using 36.1\fbinv of pp collisions data at $\roots = 13\TeV$~\cite{Aad:2019hzw}.  The measurement is performed with \ttbar candidate events in the \emu channel, and benefits from significant improvements in the calibration of the lepton identification efficiencies and the  integrated luminosity. The cross section is measured to be $\stt = 826.4 \pm 3.6 ~(\mathrm{stat}) \pm 19.6 ~(\mathrm{syst})~\mathrm{pb}$, corresponding to a total uncertainty of 2.4\%. The result is in good agreement with the theoretical prediction at NNLO+NNLL $\stt^\mathrm{th} = 832~ ^{+20}_{-29}~\mathrm{(scale)} \pm 35~(\mathrm{PDF+\as}) ~\mathrm{pb}$~\cite{Czakon:2011xx}, that has a total uncertainty of about 5\%. The measured \stt is  used to extract the value of the top quark pole mass (\mtp) by comparing to the NNLO+NNLL prediction (left-hand plot in Figure~\ref{fig}),  resulting in $\mtp = 173.1~ ^{+2.0}_{-2.1}\GeV$. Here, the precision of the result is limited by QCD scale and PDF uncertainties.  In the procedure, the dependence of the measured cross section on \mtmc is taken into account by assuming $\mtp = \mtmc$, as shown in Figure~\ref{fig} (left).  The result for \stt has also been used to estimate cross section ratios at different centre-of-mass energies, which can be used to probe the validity of existing PDF sets.

A different approach is followed by the CMS Collaboration in a measurement where \stt is measured simultaneously with  \mtmc via a profile maximum-likelihood fit to multi-differential distributions~\cite{Sirunyan:2018goh}.  The analysis is performed with 35.9\fbinv of pp collision data at $\roots = 13\TeV$, using \ttbar candidate events in the \emu channel. The cross section is measured to be $\stt = 815 \pm 2 ~(\mathrm{stat}) \pm 35 ~(\mathrm{syst})~\mathrm{pb}$,  in good agreement with the NNLO+NNLL	 theoretical prediction and the ATLAS result. The measured \stt is used to separately extract the value of \as and \mt by comparing to NNLO predictions in the \msbar scheme, using different sets of PDFs. The values of \mt and  \as are determined with a precision of 1.2\% and about 2\%, respectively. However, it has to be noted that the values of \mt and \as cannot be determined simultaneously from a measurement of \stt.

A subsequent analysis by the CMS Collaboration was designed to determine simultaneously the values of \as, \mt, and the PDFs from a triple-differential measurement of \stt. The analysis is performed with 35.9\fbinv of collisions data at $\roots = 13\TeV$, using \ttbar candidate events in the dileptonic channels~\cite{Sirunyan:2019zvx}. In this measurement, the invariant mass of the \ttbar system is chosen for is sensitivity to \mt near the production  threshold, the number of jets for its sensitivity to the value of \as, while the rapidity of the \ttbar system carries additional information on the PDFs.  The sensitivity of the analysis on the assumption of \mtmc is mitigated by means of a dedicated kinematic reconstruction algorithm that does not depend on any assumption on the value of \mt. A QCD analysis is then performed at NLO by combining the measured multi-differential cross section  with HERA deep inelastic scattering data. The inclusion of \ttbar data allows for a significant reduction of the uncertainty in the gluon PDF at momentum fractions on the order of $x_\mathrm{B} = 0.1$. Moreover, this method allows the correlation between the gluon PDF and the value of \as to be significantly reduced compared to standard PDF determinations. The values of \as and \mt are also determined with remarkable precision: The top quark pole mass is measured to be $\mtp = 170.5 \pm 0.8 \GeV$, the most precise result, to date. It is however important to note that threshold effects that are not considered in this analysis could have a significant impact on the result, as shown by a dedicated study of next-to-leading power (NLP) Coulomb corrections~\cite{Ju:2020otc}. 

Similarly, the ATLAS Collaboration has performed a PDF global fit at NNLO in QCD with NLO electroweak corrections, including HERA data in combination with W and Z cross section measurements at $\roots = 7\TeV$ and \ttbar observables measured at $\roots = 8\TeV$. Similarly to the CMS analysis,  a reduction of the PDF uncertainty at high momentum fraction is achieved~\cite{ATLAS:2018owm}.

The sensitivity of the \ttbar production threshold to \mt is exploited  by the ATLAS Collaboration in a measurement where the top quark mass is determined at NLO in the pole and \msbar schemes. The analysis is performed with 20.2\fbinv of pp collisions data at $\roots = 8\TeV$, using \ttbar candidate events in the lepton+jets final state~\cite{Aad:2019mkw}.  Here, the sensitivity is further increased by considering the invariant mass of the \ttbar $+1$ jet system, resulting in $\mtp = 171.1 \pm 1.0 ~\mathrm{(exp)} ~ ^{+0.7}_{-0.3}~\mathrm{(th)} \GeV$ and $\mtmt = 162.9 \pm 1.1 ~\mathrm{(exp)} ~ ^{+2.1}_{-1.2}~\mathrm{(th)} \GeV$ for the top-quark pole and \msbar masses, respectively. The larger theoretical uncertainty in the value of \mtmt reflects the increased dependence on the choice of QCD scales in the \msbar scheme near the production threshold.

In the \msbar scheme, the value of the top quark mass $\mt(\mu_m)$ depends on the scale $\mu_m$ at which it is evaluated. Similarly to \as, this is effect is commonly referred to as the running of \mt, and can be probed by extracting the value of $\mt(\mu_m)$ at different energy scales. This has been investigated for the first time by the CMS Collaboration, where the running of \mt is extracted from a differential measurement of \stt as a function of the invariant mass of the \ttbar system~\cite{Sirunyan:2019jyn}. The analysis is performed using 35.9\fbinv of pp collisions data at $\roots = 13\TeV$ with \ttbar candidate events in the \emu channel. The cross section is unfolded to the parton level by means of a maximum-likelihood method, in which the dependence on \mtmc is mitigated and taken into account. The extraction of the running is then performed using NLO calculations where the value of $\mu_m$ is either kept constant or set to the relevant energy scale,  following the method discussed in Ref.~\cite{Catani:2020tko}. The results are shown in Figure~\ref{fig} (right). In both cases a hypothetical no-running scenario is excluded at above 95\% confidence level.  The result can be further improved by making use of the newly available predictions at NNLO in the \msbar scheme~\cite{Catani:2020tko}.

\section{Extraction of electroweak parameters}

Electroweak contributions to the process of \ttbar production, such as the exchange of a massive neutral boson between the two final-state top quarks,  introduce a dependence of the \ttbar production cross section on fundamental electroweak parameters such as the Yukawa coupling between the top quark and the Higgs boson.  The impact of electroweak corrections is particularly relevant in the proximity of the \ttbar production threshold. This is exploited by a recent measurement by the CMS Collaboration, performed with 137\fbinv of pp collisions data at $\roots = 13\TeV$, where top-Higgs Yukawa coupling is extracted from a measurement of multi-differential distributions of dileptonic \ttbar events~\cite{Sirunyan:2020eds}. Proxy variables for the invariant mass of the \ttbar system and the rapidity difference between the top and antitop quark are built using reconstructed leptons and b-tagged jets, in order to avoid any possible bias introduced by the kinematic reconstruction of the \ttbar system.  The Yukawa coupling is then extracted by comparing to NLO MC predictions to which electroweak corrections are applied. The ratio between the measured Yukawa coupling and the standard model expectation is found to be $1.16~^{+0.24}_{-0.35}$, achieving comparable sensitivity to direct measurements with on-shell Higgs bosons.

Finally, the production cross section of a single top quark in the $t$-channel has been used by the CMS Collaboration to perform the first direct and model-independent measurement of elements of the Cabibbo-Kobayashi-Maskawa (CKM) matrix. The analysis, performed using 35.9\fbinv of pp collisions data at $\roots = 13\TeV$,  exploits multivariate techniques in order to increase the sensitivity to the values of $|V_\mathrm{tb}|$ and $|V_\mathrm{td}|^2+|V_\mathrm{ts}|^2$~\cite{Sirunyan:2020xoq}. Different scenarios of physics beyond the standard model are considered, and the total decay width of the top quark is also constrained. In this analysis, the precision in the value of $|V_\mathrm{tb}|$ is improved by about 50\% compared to a previous similar measurement by the CMS Collaboration.

\section{Summary}

Measurements of top-quark cross sections can be used to extract parameters of the QCD Lagrangian, such as \as and \mt, but also of the electroweak sector, such as the top-Higgs Yukawa coupling and elements of the CKM matrix.  The same measurements can also be used to probe the PDFs of the proton at large momentum fractions. In this proceeding, the most recent results by the ATLAS and CMS Collaborations have been summarized and discussed. Significant progress has been made with the use of LHC Run2 data at $\roots = 13\TeV$, and the increasing availability of theoretical predictions at higher orders in perturbation theory offers prospects of significant improvements in the near future.

\end{document}